# X-ray Fluorescence Sectioning

Wenxiang Cong and Ge Wang

Biomedical Imaging Division, School of Biomedical Engineering and Sciences

Virginia Tech, Blacksburg, VA 24061

**Abstract:** In this paper, we propose an x-ray fluorescence imaging system for elemental analysis. The key idea is what we call "x-ray fluorescence sectioning". Specifically, a slit collimator in front of an x-ray tube is used to shape x-rays into a fan-beam to illuminate a planar section of an object. Then, relevant elements such as gold nanoparticles on the fan-beam plane are excited to generate x-ray fluorescence signals. One or more 2D spectral detectors are placed to face the fan-beam plane and directly measure x-ray fluorescence data. Detector elements are so collimated that each element only sees a unique area element on the fan-beam plane and records the x-ray fluorescence signal accordingly. The measured 2D x-ray fluorescence data can be refined in reference to the attenuation characteristics of the object and the divergence of the beam for accurate elemental mapping. This x-ray fluorescence sectioning system promises fast fluorescence tomographic imaging without a complex inverse procedure. The design can be adapted in various ways, such as with the use of a larger detector size to improve the signal to noise ratio. In this case, the detector(s) can be shifted multiple times for image deblurring.

## 1. Introduction

Molecular imaging is sufficiently sensitive and specific for detection and quantification of biomarkers. It can be used to monitor physiological and pathological activities in an *in vivo* model at the cellular and molecular levels, such as to visualize tumour growth and therapeautic response [1]. Progress in molecular and cellular imaging depends on development of novel reporter strategies and advanced imaging technologies. Nanoparticles have been extensively studied for bioassay, biomedical imaging, as well as targeted drug and gene delivery. In living tissues, these particles are less hindered by biological barriers than larger particles of the same chemical composition, and can be better distributed [2]. Both the size and surface properties of nanoparticles are important for their interaction with biological systems.

Gold nanoparticles (GNPs) are a most important type of nanoparticles. GNPs stay in the blood longer, and repeated administration gives a higher concentration. GNPs are non-toxic and have low viscosity even at quite high concentration. Gold with the K-edge at 80.7keV enables imaging deeply inside bone and soft tissues. Also, GNPs can circulate through vessels, leak into tumors, and stay within compartments of physiological and pathological interest, because they are smaller than typical pores in the vasculature. Most interestingly, surfaces of nanoparticles including GNPs may be modified to target cells for cancer diagnosis and therapy.

X-ray fluorescence computed tomography (XFCT) allows for reconstruction of an elemental distribution within a sample from measurement of fluorescence x-rays produced by x-ray irradiation upon the sample[3,4]. The sample is excited with an x-ray energy level greater than the K-shell binding energy of an element of interest. These x-rays undergo photoelectric interaction with the atoms in the sample, resulting in the emission of fluorescence x-rays at certain characteristic energies. These resultant x-rays are detected by an energy-discriminating detector. By scanning and rotating the sample or object, it is feasible to acquire sufficient data for tomographic reconstruction of the various elements of interest in a cross-section or a volume[3].

It is highly desirable to develop a cost-effective technology for imaging of GNPs in human and animal models. With tomographic imaging of GNPs *in vivo*, one may study pathological change and therapeutic response in terms of location and concentration of GNPs in a living tissue. In 2010, Cheong et al. demonstrated the feasibility of x-ray fluorescence computed tomography (XFCT) of a small-animal-sized object containing gold nanoparticles (GNPs) at relatively low concentrations using polychromatic diagnostic x-rays [5]. Recently, Bazalova et al reported a comprehensive Monte Carlo study on XFCT and K-edge imaging [6], and showed that the contrast-to-noise ratio (CNR) of XFCT images was higher than that of transmission K-edge images for contrast concentrations below 0.4% in phantom experiments.

With the recent development of energy-discriminative photon-counting detectors, characteristic x-ray fluorescence photons can be accurately detected. The photon-counting detector can record photons of different energies separately, and have a higher signal-to-noise ratio (SNR) due to the availability of spectral information. For example, Medipix is a series of state-of-the-art photon-counting detectors. The Medipix2 chip is a pixel detector readout chip consisting of $256 \times 256$ identical elements, each working in single photon counting mode for positive or negative input charge signals [7]. It is capable of operating in 8 different energy bins.

The Medipix3 prototype chip implemented successfully a novel architecture aimed at eliminating the spectral distortion produced by the charge sharing process [8]. The Medipix design allows use of various detection substrates including Si, GaAs, and CdTe for preclinical and clinical x-ray studies.

In this paper, we propose a new x-ray fluorescence imaging system for elemental analysis. This system works in the fan-beam excitment mode, collects fluorescence data across any plane defined by the irradiating fan-beam, and measures the resultant elemental distribution directly without a complex inverse procedure. Our design could take a detector of large size to improve the signal to noise ratio. In that case, the detector array(s) can be shifted multiple steps for image deblurring. In the following section, we will first propose a novel x-ray fluorescence imaging system and then give a analytic formula for computation of the elemental concentration. In the third section, contrast resolution analysis and Compton scattering correction are described. In the last section, relevant issues are discussed, and the paper is concluded.

## 2. Methodolgy

### 2.1. Imaging system for direct elemental mapping

The proposed x-ray fluorescence imaging system consists of a common x-ray source (ISOVOLT Titan E Series, 160kVp/19mA, GE), an x-ray filter in front of the source to absorb low energy x-rays for a narrow spectral range suitable for K-edge imaging. A slit collimator is then used to shape x-rays into a fan-beam to illuminate an object. Relevant elements such as gold nanoparticles on the fan-beam plane in an object are excited to generate x-ray fluorescence signals, as shown in elemental distribution on the fan-beam plane in reference to the attenuation properties of the object and the imaging parameters associated with the fan-beam. The new imaging mode does not request any scanning for multiple projection angles, and can achieve a fast tomographic imaging performance.

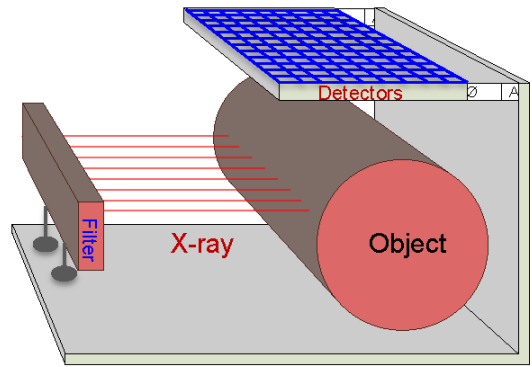

**Figure 1.** X-ray fluorescence sectioning system.

Figure 1. One or more 2D spectral detectors are placed to face the fan-beam plane to measure x-ray fluorescence data. The anti-scatter collimators are needed to block the x-rays that do not come from targeted locations on the fan-beam plane. In other words, the collimators are installed in front of the corresponding detector elements so that each detector element only see a unique area element on the irradiated plane and just record the x-ray fluorescence signal from that area element. The measured 2D x-ray fluorescence data can be processed to obtain the targeted

When an incident x-ray fan beam penetrates a biological tissue, the x-ray intensity (photons cm$^{-2}$ s$^{-1}$) distribution can be computed from linear attenuation coefficients using Beer-Lambert's law:

$$I(x, y, z_{slic}) = I_0 \exp\left(-\int_{L(x,y,z_{slic})} \mu_t(w) dw\right) \tag{1}$$

where the fan beam plane is defined as $\{(x, y, z_{slic}) | -\infty < x < \infty, -\infty < y < \infty, z_{slic} = \text{constant}\}$, $I_0$ is the source intensity, $\mu_t$ is the attenuation coefficient (cm$^{-1}$) distribution in the object, $L(x, y, z_{slic})$ is a line segment from the source at $(x_0, y_0, z_{slic})$ to a point $(x, y, z_{slic})$ in the object. After a primary x-ray beam of a proper energy level interacts with targeted elements in the object, characteristic x-rays are emitted with the intensity proportional to the product of the concentration of targeted elements $N(x, y, z_{slic})$ ($\mu g/ml$), the photoelectric mass absorption coefficient of the targeted elements $\mu_{ph}$ (cm$^2$/g), and the fluorescence yield $\eta$. The intensity of the fluorescence x-rays measured by the detector can be formulated as

$$P(x, y, z_{slic}) = \eta \mu_{ph} I(x, y, z_{slic}) N(x, y, z_{slic}) \cdot \text{Ve} \cdot \exp\left(-\int_0^d \mu_F(x, y, z_{slic} + z) dz\right) \frac{A_D}{4\pi d^2} \tag{2}$$

where $A_D$ is the area of detector element (cm$^2$), and $d$ (cm) is the distance between detector element and a point on the fan-beam plane, and Ve is the volumetric element, which is the intersection of the x-ray fan-beam and the projected area element on the fan beam plane as seen by the detector. From the measured fluorescence intensity in Eq. (2), the concentration of the targeted elements $N(x, y, z_{slic})$ can be computed directly. Eq. (2) is an analytic formula for the concentration of targeted elements from the fluorescence intensity distribution measured by the x-ray photon counting detector.

## 2.2. Imaging system for deblurring an elemental map

This alternative system design allows the use of a photon counting detector of large-size element. In this case, the imaging system can achieve a higher signal-to-noise ratio with a collimator to further suppress scattered radiation and fluorescence crosstalk. To achieve a higher spatial resolution, the detector array(s) can be shifted multiple steps along transverse and longitude directions respectively for image deblurring. A linear relationship can be established between the measured data and the fluorescence intensity distribution on a refined sampling grid. By solving the linear system of equations, we improve the spatial resolution of the target element distribution in the object.

Without loss of generality, here we present a special case in which the spectral detector element size is the four times of spatial resolution to be achieved. In this case, we acquire four datasets by shifting the spectral detector array three times as shown in Figure 2. The photon counting detector array (with a $\frac{n}{2} \times \frac{m}{2}$ detector elements) covers a region of interest in the object to acquire the first dataset. Then, the spectral detector array is shifted by a spatial resolution size transversely to acquire the second dataset. The spectral detector array at the position for the first dataset is shifted with a spatial resolution size longitudinally to acquire the third dataset. Finally, the spectral detector array is further shifted by a spatial resolution size transversely to acquire the fourth dataset. From these four datasets, we have a linear system of equations with respect to the fluorescence intensity distribution on the refined sampling grid:

$$\begin{cases} x_{i,j} + x_{i,j+1} + x_{i+1,j} + x_{i+1,j+1} = I_{i,j}^1, & i=1,2,\cdots,n-1,\ j=1,2,\cdots,m-1 \\ x_{i,j} + x_{i,j+1} + x_{i+1,j} + x_{i+1,j+1} = I_{i,j}^2, & i=1,2,\cdots,n-1,\ j=2,2,\cdots,m \\ x_{i,j} + x_{i,j+1} + x_{i+1,j} + x_{i+1,j+1} = I_{i,j}^3, & i=2,2,\cdots,n,\ j=1,2,\cdots,m-1 \\ x_{i,j} + x_{i,j+1} + x_{i+1,j} + x_{i+1,j+1} = I_{i,j}^4, & i=2,2,\cdots,n,\ j=2,2,\cdots,m \end{cases} \quad (4)$$

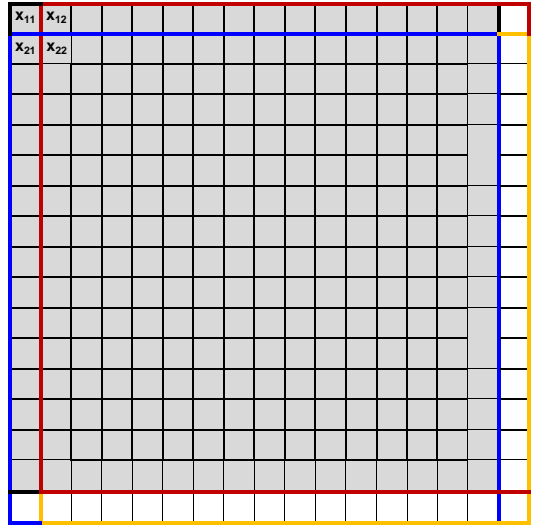

**Figure 2.** Schematic diagram of a spectral detector array.

With the zero boundary conditions:

$$x_{n+1,j} = 0, \quad x_{i,m+1} = 0, \quad i=1,2,\cdots,n+1,\ j=1,2,\cdots,m+1 \quad (5)$$

Depending on applications, natural boundary conditions can be also applied to solve the linear system. The linear system of equations (4)-(5) is well-posed, and has a unique and stable solution with respect to the fluorescence indensity distribution on a refined sampling grid (m×n). Hence, the distribution of the targeted elements in the object can be computed using Eqs. (3)-(5) for improved spatial resolution.

**2.3. Numerical simulation:** A phantom was employed to evaluate the above-proposed imaging system design. The phantom was of 64×64mm$^2$, and consisted of 40 disks in the background. Each disk was assigned with a different fluorescence density in the range of [0.1×10$^{-6}$, 0.6×10$^{-6}$] μW, as shown in Fig. 3(a). We performed four times measurements of fluorescnece data using a 32×32 photon-counting detector array, each of whose elements covered four neighboring pixels in the phantom image. The four measured images were corrupted by Poisson noise to mimic a real measurement condition, as shown in Figure 3(b-e). Then, the image deblurring method was performed to reconstruct the fluorescence intensity distribution. The reconstructed result was in excellent agreement with the truth, as shown in Figure 3(f). This example clearly demonstrates the feasibility of our x-ray fluorescence sectioning mode.

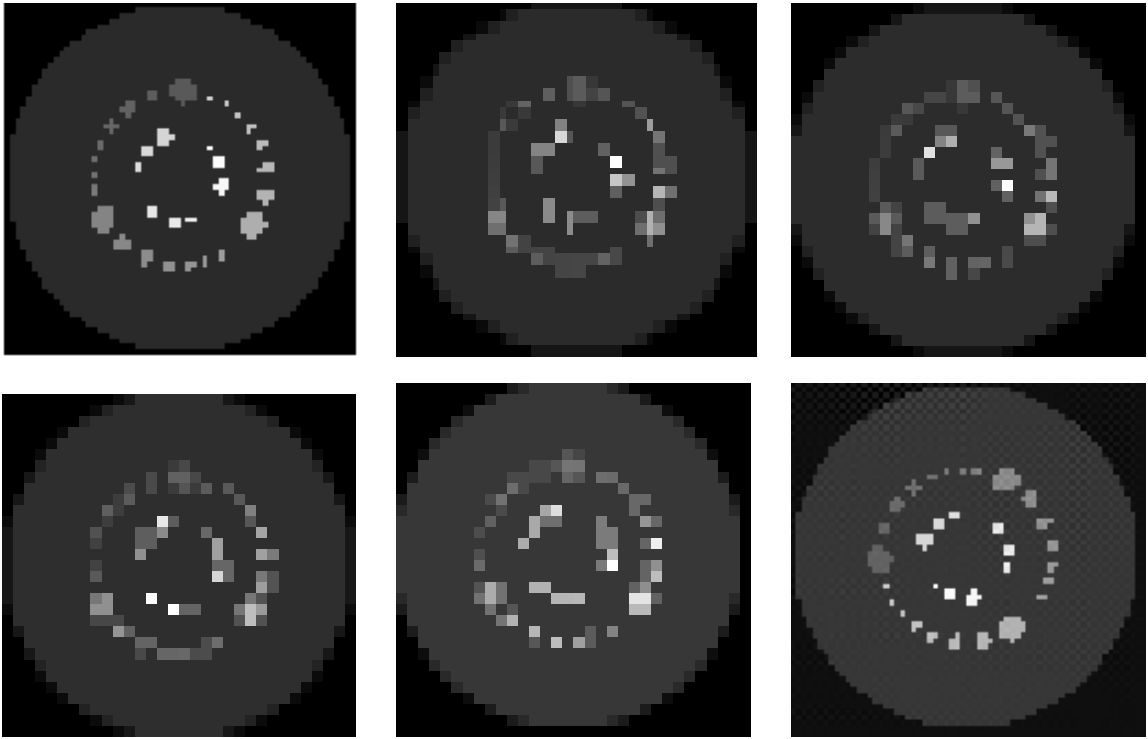

**Figure 3.** X-ray fluorescence sectioning through a numerical phantom. (a) The true fluorescence intensity image, (b-e) measured fluorescence intensity images in a low resolution, and (f) the reconstructed fluorescence intensity image in an improved resolution.

## 3. Contrast resolution of x-ray fluorescence sectioning

Let take the gold nanoparticle (GNP) as an example to analysize the signal to noise ratio and contrast resolution of x-ray fluorescence sectioning. The photoelectric mass absorption coefficient of gold is 1.72cm$^2$/g at 80 keV. Gold K-edge is at 80.7 keV. While an x-ray beam goes through a biological tissue to excite GNPs, x-ray photons may induce photoelectric interactions with gold atoms, and generate isotropically distributed fluorescence x-ray photons. Relevant fluorescent lines of gold are at 67.0 and 68.8 keV corresponding to Au Kα2 and Kα1, respectively. The photon-counting detector can be operatated in the spectral bin [63, 73] keV to extract fluorescence signals, consistent to the energy resolution of Medipix-3. X-ray fluorescence has high yields of 100% at Ka1 and 58.6% at Ka2 peaks and strong penetrability through the sample. At each area element on the fan-beam plane, gold fluorescence photons were emitted proportional to the concentration of Au. We assumed that the concentration of GNPs in a sample was 10-100μg/ml for analysis of contrast resolution, which is biologically meaningful, the source to object distance was 2.6cm, the object diameter was 3cm, and the linear attenuation coefficient of the background was 0.2 cm$^{-1}$ to mimic biological tissues. We use an x-ray source operated at 160kVp/19mA, which has a about 10% effective x-ray energy to irradiate targeted nanoparticles. The number of fluorescence photons per second detected by a detector element can be estimated roughly based on Eq. (2) as follows:

$$P(x, y, z_{slic}) = \eta\mu_{ph}I(x, y, z_{slic})N(x, y, z_{slic})Ve \cdot \exp\left(-\int_0^d \mu_F(x, y, z_{slic} + z)dz\right)\frac{A_D}{4\pi d^2}$$

$$\geq 0.586 \times (1.72 cm^2/g) \times \left(\frac{0.1 \times 160 keV \times 19mA}{4\pi \times 2.6^2 cm^2}/s \times \exp(-3 \times 0.2)\right) \times (10^{-5} g/cm^3) \quad (6)$$

$$\times (10^{-3} cm^3) \times (\exp(-3 \times 0.2)) \times \frac{1mm^2}{4\pi \times 30^2 mm^2}$$

$$= 9.6 \times 10^{-13} mW/s \approx 80\, photons/s$$

For GNP excitation, the x-ray spectrum must be significant above 70keV. The x-rays beyond this energy level also induces Compton scattering in the object. The detector array is placed parallel to the incident beam direction. Although the possibility of Compton scattering photon reaching the detector is small at this case, the Compton scattering compromises the fluorescence signal. The measured signal by the detector should be the sum of Kα fluorescence, Compton scattering, and background noise. To convert the fluorescence to the concentration of gold in the object, the Compton scattering and background noise must be filtered out.

Here we develop a method to separate the fluorescence signal from the Compton scattering background in the fluorescence spectral window. The Klein-Nishina cross section gives a probability of a photon scattering into a given angle [9]:

$$\frac{d\sigma_{compton}}{d\Omega}(\beta) = \frac{r_e^2}{2[1+\alpha(1-\cos\beta)]^2}\left[(1+\cos^2\beta) + \frac{\alpha^2(1-\cos\beta)^2}{1+\alpha(1-\cos\beta)}\right] \quad (7)$$

where $\beta$ is the scattering angle, $\alpha = E_\gamma/m_e c^2$, $E_\gamma$ is the photon energy, $m_e$ is the electron mass, $c$ is the speed of light, and $r_e$ is the classical radius of the electron. The Compton scattering is closely related to the electron density distribution. The linear attenuation coefficient of background medium can be reconstructed with computed tomography, and the electron density distribution can be estimated. The Compton scattering intensity in the fluorescence spectral window mearsured by the detector can be computed as follows

$$C(x,y) = I(x,y)\rho(x,y)\frac{d\sigma_{compton}}{d\Omega}(\beta)\exp\left(-\int_{Le}\mu(s)ds\right)A_s A_D/d^2 \quad (8)$$

where $\beta$ is the scattering angle, and $\rho(x,y)$ is the electron density, $A_s$ is an area element on the fan-beam plane. From Eq. (8), Compton scattering from the area element on the fan-beam plane reaching the detector element can be predicted. Furthermore, the Compton scattering signal from the targeted elements at angles within the the detector aperture can be also separated from fluorescence data as follows:

$$P(x,y,z_{slic}) - C_b(x,y) = N(x,y,z_{slic})\left(\eta\mu_{ph}I(x,y,z_{slic})Ve \cdot \exp\left(-\int_0^d \mu_F(x,y,z_{slic}+z)dz\right)A_D/d^2\right) \quad (9)$$

**4. Discussions and conclusion**

The proposed x-ray fluorescence sectioning concept is based on the advanced photon-counting detector technology and in a direct observation mode. As a result, both the signal to noise ratio and imaging speed are improved. The photon-counting detector can record different photon energies at the same time, and provide both high energy resolution and spatial resolution. Photon-counting detectors have an inherently higher signal-to-noise ratio (SNR) because electronic noise below the counting threshold is eliminated. The performances of x-ray fluorescence sectioning can be significantly enhanced by incorporating the state of the art photon-counting techniques.

We have computed the number of fluorescence photons detected by an elemental detector per second for GNPs based on Eq. (2), which shows that sufficiently many photons can reach the detector in preclinical applications. Since over 70 keV x-ray Compton scattering would dominate the photon interaction with matter, an anti-scatter grid is important to effectively elimate the scattering outside the detector aperture. The Compton scattering within the detector aperture is predicted by Eq. (8), and can be substracted from measured data to enhance the x-ray fluorescence signal to noise ratio, allowing an accurate quantification of target elements in the object. Moreover, Monte Carlo (MC) simulation can be used to give a more accurate estimate of the Compton scattering background.

This system-ray fluorescence sectioning modes can reveal the elemental composition and support novel contrast-enhanced studies with functional, cellular and molecular imaging. This imaging approach promises unique imaging performance especially in terms of temporal consistence and energy sensitivity.

In conclusion, we have proposed a new x-ray fluorescence imaging architecture for elemental analysis. This approach allows fast fluorescence tomographic imaging for accurate reconstruction of one or multiple nanoparticle distributions in any cross-section without a complex inverse procedure. The new imaging mode does not request the rotation of either an object or a detector, suggesting important preclinical or biological applications.

**Acknowledgements**

The work has been partially supported by NIH R01HL098912.